\documentclass[%
 aip,
 amsmath,amssymb,
 reprint,%
]{revtex4-2}

\usepackage{graphicx}
\usepackage{dcolumn}
\usepackage{bm}

\usepackage[utf8]{inputenc}
\usepackage[T1]{fontenc}
\usepackage{mathptmx}
\usepackage{etoolbox}
\usepackage{float}

\makeatletter
\def\@email#1#2{%
 \endgroup
 \patchcmd{\titleblock@produce}
  {\frontmatter@RRAPformat}
  {\frontmatter@RRAPformat{\produce@RRAP{*#1\href{mailto:#2}{#2}}}\frontmatter@RRAPformat}
  {}{}
}%
\makeatother
\begin{document}
\title{One-sided composite cavity on an optical nanofiber for cavity QED}
\author{Ramachandrarao Yalla*}
\email{rrysp@uohyd.ac.in}
\affiliation{School of Physics, University of Hyderabad, Hyderabad, Telangana, India-500046}
\author{K. Muhammed Shafi}
\altaffiliation{Department of Instrumentation \& Applied Physics, Indian Institute of Science, Bengaluru, India-560012}
\author{Kali P. Nayak}
\altaffiliation{Department of Engineering Science, University of Electro-Communications, Chofu, Tokyo, Japan-182-8585}
\author{Kohzo Hakuta}
\affiliation{Center for Photonic Innovations, University of Electro-Communications, Chofu, Tokyo, Japan-182-8585}
\date{\today}

\begin{abstract}
We demonstrate a one-sided cavity on an optical nanofiber (ONF) using a composite method. The one-sided composite cavity is created by mounting an asymmetric defect mode grating on an ONF. We design the one-sided composite cavity on an ONF to enhance channeling efficiency into one side of ONF while operating from under- to critical- and over-coupling regimes using numerical simulations. Experimentally, we demonstrate  coupling characteristics of the one-sided composite cavity, showing good correspondence with simulation results. 
\end{abstract}

\maketitle
Cavity quantum electrodynamics (QED) approach for enhancing light-matter interaction strength has attracted great interest with potential applications ranging from quantum optics\cite{Kimble,Ritter,Yalla14,Kali19}, quantum networks\cite{Kimble,Ritter}, and sensing\cite{Yang}. A crucial aspect of cavity QED is Purcell enhancement and unidirectional channeling of spontaneous emission of single quantum emitters to realize deterministic single photon sources\cite{Haroche,Thompson, Fushman, Tiecke,Englund}. Two types of cavity schemes have been discussed, such as two-sided and one-sided \cite{Milburn}. Formation of two/one-sided cavity schemes has been proposed and experimentally demonstrated in various geometries at micro/nano-scales\cite{Reiserer,Vahala,Liu,Aoki,Meschede, Reiserer}. Examples include conventional Fabry-Perot, diamond nanobeam\cite{Thompson,Tiecke}, microtoroid and bottle-neck \cite{Aoki2006, Junge}, silicon nitride photonic crystal (PhC) \cite{Thompson,SPYu}, and silica optical micro/nano-fiber based cavities\cite{Sumetskya,Kali11,KaliPhC1,Schell15,SNC1,Limin,Philipp}. The formed cavities have been utilized for enhancing the spontaneous emission rate of single quantum emitters\cite{Reiserer,Vahala,Liu,Aoki,Meschede}. 

Among the examples mentioned above, silica PhC optical nanofiber (ONF) cavities are particularly promising due to their ultra-low-loss coupling to conventional single-mode fibers for fiber-based quantum networks\cite{Review18,Yalla12}. Regarding the formation of two-sided PhC cavities on silica ONFs, two methods have been proposed and experimentally demonstrated\cite{KaliPhC1,Schell15,SNC1,Mark13,Jam15,Takashima, Li,Tashima,Romagnoli}. One is the direct fabrication of nano-structures on the surface of ONF itself using different techniques \cite{Kali11,KaliPhC1,Schell15,SNC1}. The other is a composite photonic crystal cavity (CPCC) method, wherein CPCC is created by combining a defect mode grating with an ONF \cite{Mark13,Yalla14,Jam15}. CPCC method is particularly preferable for positioning solid-state quantum emitter to anti-node of cavity mode\cite{Yalla14}. Precise tuning of cavity resonance wavelength up to $\pm$10 nm around the designed wavelength to match quantum emitter's spectral emission line has also been experimentally demonstrated\cite{Schell15,Yalla20}. PhC ONF cavities have been utilized for enhancing the spontaneous emission rate of single quantum emitters in the vicinity of an ONF \cite{Yalla14, Kali19,Schell15}. These PhC ONF cavities are two-sided, and single photons are transmitted equally from both sides of the cavity on an ONF i.e. both sides of ONF guided modes. The measured maximum channeling efficiency into either side of ONF guided mode was $32$-$42$\% \cite{Yalla14,Kali19}. 

A one-sided cavity will be crucial for practical applications based on single photon sources. The one-sided cavity can channel the total spontaneous emission of a single quantum emitter into one side of ONF guided modes. Moreover, various quantum information protocols are formulated based on one-sided cavity schemes \cite{Reiserer}. Therefore, realizing one-sided cavities on ONFs is an essential requirement to extend for various quantum information applications. The one-sided PhC cavity on ONFs has not been experimentally demonstrated yet.

\begin{figure*}[ht]
\centering
\includegraphics[width=15cm]{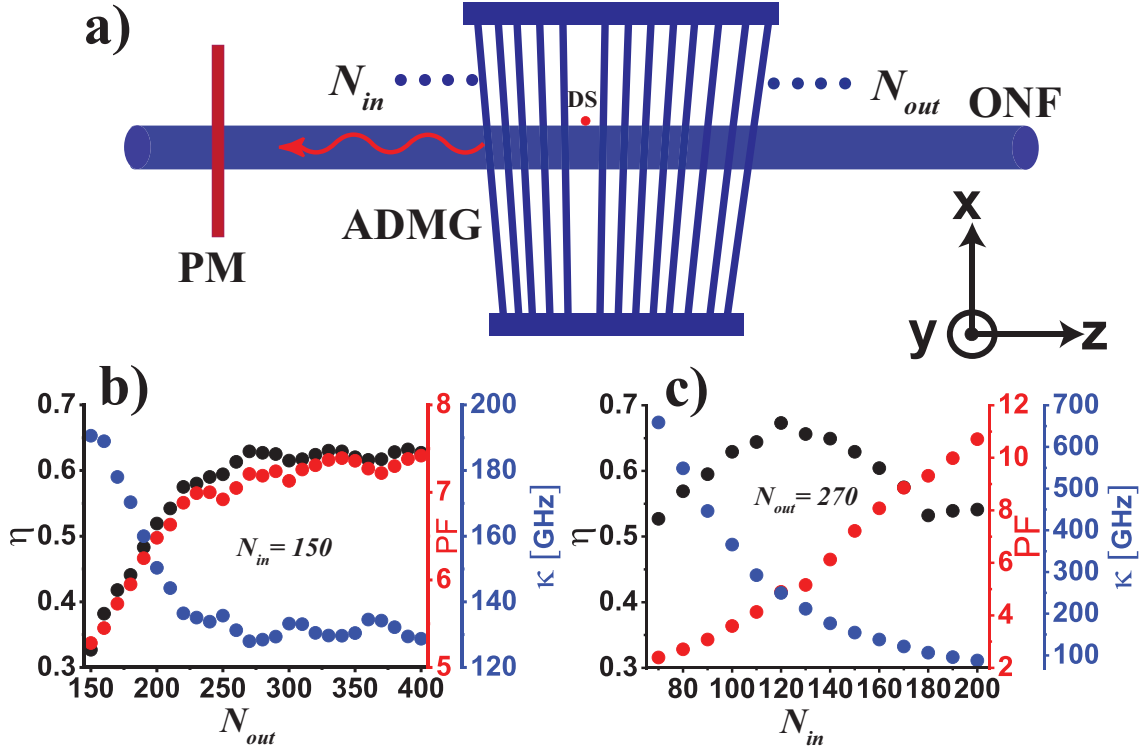}
\caption{{\label{Conceptual diagram}} (a) A conceptual top view of a one-sided composite cavity on an optical nanofiber (ONF). The one-sided composite cavity is formed by combining an ONF and an asymmetric defect mode grating (ADMG). $N_{in}$, $N_{out}$, PM, and DS denote input slat number, output slat number, power monitor, and dipole source, respectively. (b) and (c) show simulation predicted values for  channeling efficiency ($\eta$, black dots), Purcell factor ($PF$, red dots), and total cavity linewidth ($\kappa$, blue dots) as a function of $N_{out}$ and $N_{in}$-values, respectively. $N_{in}$-value is fixed at $150$ for (b) and $N_{out}$-value is fixed at $270$ for (c).} 
\end{figure*}

In this letter, we report a systematic design and implementation of a one-sided composite cavity on an ONF. As conceptually displayed in Fig.~\ref{Conceptual diagram}(a), the essential point is to fabricate a nano-grating with a different number of slats on either side of the defect, defined as an asymmetric defect-mode grating (ADMG). Then the one-sided composite cavity is formed by mounting ADMG on ONF. We numerically design the composite cavity to enhance channeling efficiency into one side of ONF while operating  from under- to critical- and over-coupling regimes with a minimum scattering loss by a varying number of slats at the input side of the cavity. Experimentally, we demonstrate coupling characteristics of the one-sided composite cavity, showing good correspondence with simulation results.

First, we discuss the design of a one-sided composite cavity. The conceptual diagram of a one-sided composite cavity on ONF is shown in Fig.~\ref{Conceptual diagram}(a). The cavity is formed by mounting ADMG onto ONF. The essential point is to design a nano-grating with different number of slats on either side of the defect. Additionally, we designed the ADMG with a diverged grating period for tuning the cavity resonance\cite{Yalla20}. Simulations are performed using the finite difference time domain method\cite{Yalla20}. We set a $\it{y}$-polarized dipole source (DS) on the surface of ONF at the anti-node position of the cavity, and a power monitor (PM) is placed sufficiently far from the DS position to measure the power ($P_c$) coupled into the guided mode. $P$ and $P_0$ are the total power emitted by the DS in the presence of the cavity and the vacuum environment, respectively. We determine the Purcell factor ($PF$) and the channeling efficiency ($\eta$) as $PF$= $P/P_0$ and $\eta$= $P_c/P$, respectively. We find the optimum parameters by maximizing the $\eta$-value into the guided mode. We also simulate the cavity reflection spectra using a mode source to infer the total cavity linewidth ($\kappa$) and the on-resonance reflection ($R_0$).
 
Design parameters for the one-sided composite cavity are ONF diameter ($2a$), grating period ($\Lambda_g$), defect-width ($w_g$= 1.5$\Lambda_g$), duty cycle ($\alpha$), slat width ($t$= $\alpha\Lambda_g$), number of slats at the input side ($N_{in}$), and number of slats at the output side ($N_{out}$). We assume a rectangular slat shape with a slat depth of 2 $\mu$m and ADMG length to be $500$ $\mu$m \cite{Yalla20}. As per our previous work \cite{Yalla20}, optimum parameters for the two-sided cavity were as follows: $2a$= $510$ nm, $\Lambda_g$= $252$ ($\pm5$) nm, $w_g$= $378.0$ ($\pm7.5$) nm, $\alpha$= $20$\%, $t$= $50.4$ ($\pm$1.0) nm, and total number of slats ($N$)= $300$. The essential point was an equal number of slats on both sides of the defect i.e. $N_{in}$=$N_{out}$= $150$, where $N$= $N_{in}$+$N_{out}$ is total number of slats. The simulated $\eta$-value was same for both sides of ONF guided modes for two-sided cavity. The maximum $\eta$-value was $0.325$ at one-side of ONF guided modes.  

In the present design, we find optimum parameters for the one-sided composite cavity by simulating $\eta$-value, $PF$-value, and $\kappa$-value. A key point of the optimization is to choose optimum $N_{in}$ and $N_{out}$-values while keeping scattering loss as small as possible to achieve $\eta$-value as maximum as possible at one side of ONF. By monitoring $\eta$ and $PF$-values, $N_{out}$-value is swept from $150$ to $400$ with a step of $10$ while keeping $N_{in}$-value fixed at $150$. Summary of the simulated results is plotted in Fig.~\ref{Conceptual diagram}(b). Black, red, and blue dots correspond to $\eta$, $PF$, and $\kappa$-values as a function of $N_{out}$-value, respectively. One can readily see that saturation behavior of $\eta$, $PF$, and $\kappa$-values for $N_{out}$ $\ge$ $270$. $\eta$ and $PF$-values are saturated for $\kappa$$\le$ $130$ GHz. 

By setting $N_{out}$= $270$, $N_{in}$-value is swept from $70$ to $200$ with a step of $10$. Summary of the simulated results is plotted in Fig.~\ref{Conceptual diagram}(c). Black, red, and blue dots correspond to $\eta$, $PF$, and $\kappa$-values as a function of $N_{in}$-value, respectively. One can readily see that $\kappa$-value decreases as $N_{in}$-value increases; consequently, $PF$-value increases as expected. However, the $\eta$-value peaks around $N_{in}$= $120$ ($\kappa$= $250$ GHz). Maximum $\eta$-value is found to be $0.673$. Note that the present $\eta$-value is twice compared to the case for the two-sided cavity on the ONF \cite{Yalla20}. The whole emission can be channeled into one side of ONF guided modes, useful for various single-photon applications. Thus, optimum input and output slat numbers for the one-sided composite cavity are $N_{in}$= $120$ and $N_{out}$$\ge$$270$, respectively. 
Note that the tunability of the presently designed cavity is $\pm$$10$ nm.

\begin{figure}[ht]
\centering
 \includegraphics[width=9cm]{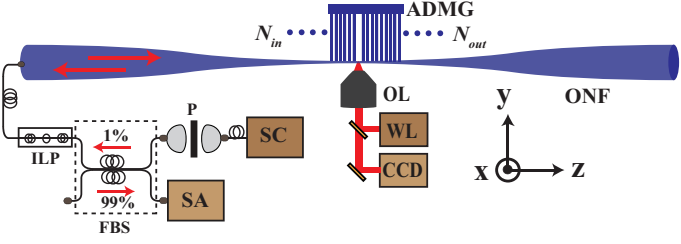}
\caption{{\label{Experimental Schematic}} A schematic of the experimental setup for measuring reflection spectra of a one-sided composite cavity. SC, P, ILP, FBS, OL, WL, CCD, and SA denote super-continuum source, polarizer, in-line polarizer, in-line fiber beam splitter, objective lens, white light source, charge-coupled device, and spectrum analyzer, respectively. $N_{in}$ and $N_{out}$ denote input slat number and output slat number of ADMG, respectively.} 
\end{figure}

Regarding experimental realizations, optimized parameters for ADMG patterns are fabricated on a silica substrate using electron beam lithography and chemical etching \cite{Mark13,Yalla14, Jam15,Yalla20}. By considering experimental imperfections, ADMG patterns are fabricated with $N_{out}$= $390$ and $N_{in}$-value varied from $100$ to $190$ with a step size of $10$. Note that patterns with different $N_{in}$-values are fabricated on the same silica substrate. The ONF is fabricated using heat and pull technique\cite{Review18,Jam17,Jam15} and has a uniform waist region of $2.5$ mm with a diameter of $510$$\pm$5 nm\cite{Review18,Jam17,Jam15}. The measured optical transmission of the ONF is $98$\%. 

A schematic of the experimental setup for optical characterization is shown in Fig. \ref{Experimental Schematic}. We image the ADMG pattern through an objective lens (OL) by sending white light (WL) and observing using a charge-coupled device (CCD). ONF is positioned perpendicular to the slats using a rotational stage. The detailed procedure for the precise alignment of the ADMG pattern and its mounting onto ONF can be found in Refs\cite{Mark13, Yalla14,Jam15}. At the input side of the cavity, we launch a spectrally filtered supercontinuum (SC) light with a wavelength ranging from $600$ to $700$ nm through a polarizer (P) to ensure linear polarization. The polarized light is split into two beams with a ratio of 1:99 using a fiber beam splitter (FBS). One of the output sides (1\% of the input signal) of the FBS is connected to the cavity through an in-line fiber polarizer (ILP) to control the light polarization angle. The resultant reflected signal is monitored through the other input side of the same FBS with the transmission of 99\% and recorded using a spectrum analyzer (SA) with a resolution of $0.05$ nm ($30$ GHz). 

\begin{figure}[hb]
\centering
\includegraphics[width=9cm]{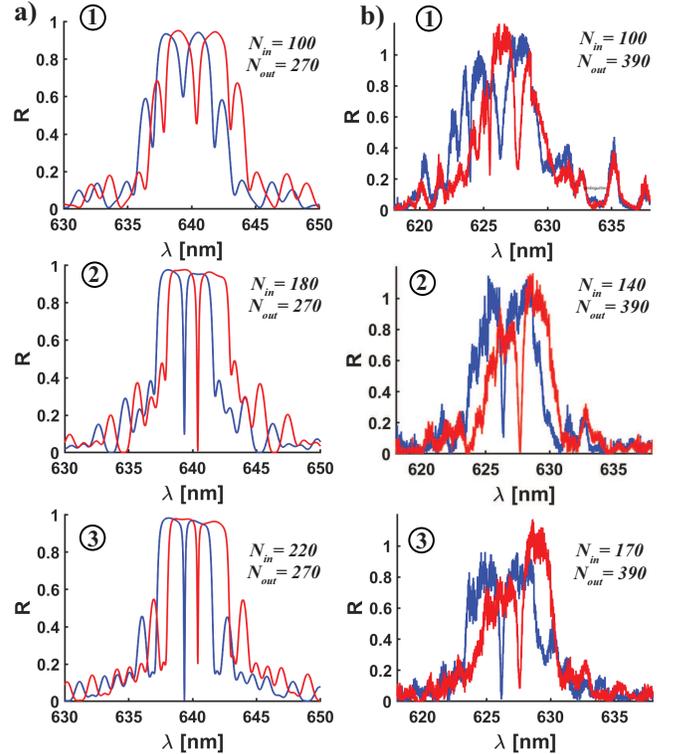}
\caption{{\label{Results1}}Typical simulated and the corresponding measured results for a one-sided composite cavity on an optical nanofiber: (a) and (b) show simulated and measured cavity reflection spectra for $x$ (blue trace) and $y$ (red trace)-modes. $1$, $2$, and $3$ correspond to different input slat number ($N_{in}$) at fixed output slat number ($N_{out}$) as mentioned in the figure.}
\end{figure}

Simulated and the corresponding measured one-sided composite cavity characteristics are shown in Figs.~\ref{Results1}(a) and (b), respectively. Typical simulated cavity reflection spectra for $N_{in}$= $100$, $180$, and $220$ are shown in Figs.~\ref{Results1}(a)-$1$, $2$, and $3$, respectively. Blue (red) trace corresponds to $x$ ($y$)-mode. One can readily see strong photonic reflection bands around $640$ nm, along with dips at the center of the bands for all traces. We obtain cavity resonance wavelength ($\lambda_0$), cavity linewidth ($\Delta\lambda$), and on-resonance reflection ($R_0$) by fitting the dip in reflection spectra with Lorentzian. The $x$- and $y$-modes resonance dips are separated by $1.05$ nm. We measured cavity reflection spectra at various $N_{in}$-values by un-mounting ADMG at one pattern and translating it to the other pattern for the next mounting. Typical measured cavity reflection spectra in Figs.~\ref{Results1}(b)-$1$, $2$, and $3$ correspond to $N_{in}$= $100$, $140$, and $170$, respectively. Note that due to experimental imperfections, $N_{in}$ and $N_{out}$-values are different compared to simulation values. Blue (red) trace corresponds to $x$ ($y$)-mode. We observed strong optical reflection bands around a wavelength of $627$ nm in all traces, accompanied by dips at the center. Measured separation in $\lambda_0$-values for $x$- and $y$-modes is $1.30$ nm. The total cavity linewidths ($\kappa$= $c\Delta\lambda/\lambda_0^2$, where $c$ is velocity of light in free space) and quality factors ($Q$= $\lambda_0/\Delta\lambda$) are obtained from $\Delta\lambda$-values and $\lambda_0$-values. Simulated and the corresponding measured $\lambda_0$, $\Delta\lambda$, $R_0$,  $Q$, and $\kappa$-values are summarized in Table~\ref{Table1}. Superscripts denote polarizations.

\begin{table*} [htb]
\caption{{\label{Table1}}Summary of typical simulated and measured parameters for the one-sided composite cavity on an optical nanofiber.}
\centering%
\begin{tabular}{|p{1.5cm}|p{0.6cm}|p{1cm}|p{1cm}|p{1cm}|p{1cm}|p{1cm}|p{1cm}|p{1cm}|p{1cm}|p{1cm}|p{1cm}|}
\hline
&{$N_{in}$} &\multicolumn{2}{c|}{$\lambda_0 [nm]$}&\multicolumn{2}{c|}{$\Delta\lambda [nm]$}&\multicolumn{2}{c|}{$R_0$}&\multicolumn{2}{c|}{$Q$}&\multicolumn{2}{c|}{$\kappa [GHz]$}\\
\cline{1-12}
&&\it{$\lambda_0^{x}$}&\it{$\lambda_0^{y}$}&\it{$\Delta\lambda^{x}$}&\it{$\Delta\lambda^{y}$}&\it{$R_0^{x}$}&\it{$R_0^{y}$}&\it{$Q^x$}&\it{$Q^y$}&\it{$\kappa^x$}&\it{$\kappa^y$}\\
\cline{1-12}
&100&639.33&640.38&0.552&0.500&0.595&0.453&1158&1280&405&365\\
\cline{2-12}
Simulation&180&639.33&640.38&0.167&0.146&0.095&0.008&3828&4386&122&107\\
\cline{2-12}
&220&639.33&640.38&0.109&0.100&0.004&0.200&5870&6404&79&73\\
\hline
&100&626.30&627.60&$0.83$&$0.61$&$0.33$ &$0.26$&$754$&$1028$&$635$ &$464$\\
&&$\pm$$0.05$ &$\pm$$0.05$ &$\pm$$0.08$ &$\pm$$0.06$&$\pm$$0.04$ &$\pm$$0.03$&$\pm$$50$ &$\pm$$90$&$\pm$$60$ &$\pm$$50$\\
\cline{2-12}
Experiment&140&626.38&627.66&$0.47$&$0.39$&$0.15$&$0.02$&$1332$&$1610$&$358$&$297$\\
&&$\pm$$0.05$ &$\pm$$0.05$ &$\pm$$0.04$ &$\pm$$0.03$&$\pm$$0.06$ &$\pm$$0.001$&$\pm$$90$ &$\pm$$120$&$\pm$$30$&$\pm$$35$\\
\cline{2-12}
&170&626.20&627.55&$0.30$&$0.32$&$0.02$&$0.08$&$2087$&$1960$&$230$&$244$\\
&&$\pm$$0.05$ &$\pm$$0.05$ &$\pm$$0.04$ &$\pm$$0.05$&$\pm$$0.01$ &$\pm$$0.01$&$\pm$$140$&$\pm$$130$&$\pm$$50$ &$\pm$$30$\\
\hline
\end{tabular}
\end{table*}

Regarding measured $\lambda_{0}$-values and separation for $x$- and $y$-modes, deviation from simulation values are mainly due to fabrication ambiguities in ONF diameter and ADMG parameters. As discussed in previous works\cite{Yalla14, Yalla20}, using measured $\lambda_{0}$-value and fluctuations in the grating period, we estimate the  ONF diameter to be $500$$\pm$$4$ nm \cite{Yalla20}, implying that the ONF diameter is thinner than the value set in simulations. The changes required in ONF diameter indicate that the present results are within fabrication tolerances.  
However, it should be noted that the presently designed cavities can tune the $\lambda_{0}$-value up to $\pm$10 nm\cite{Yalla20}. Although experimental discrepancies exist, measured reflection-band, clear polarization dependence, and cavity mode around the center of the reflection-band are in reasonable agreement with simulation predicted results. To have a quantitative evaluation of the designed cavity, we analyze coupling characteristics in the following.

\begin{figure}
\centering
\includegraphics[width=8 cm]{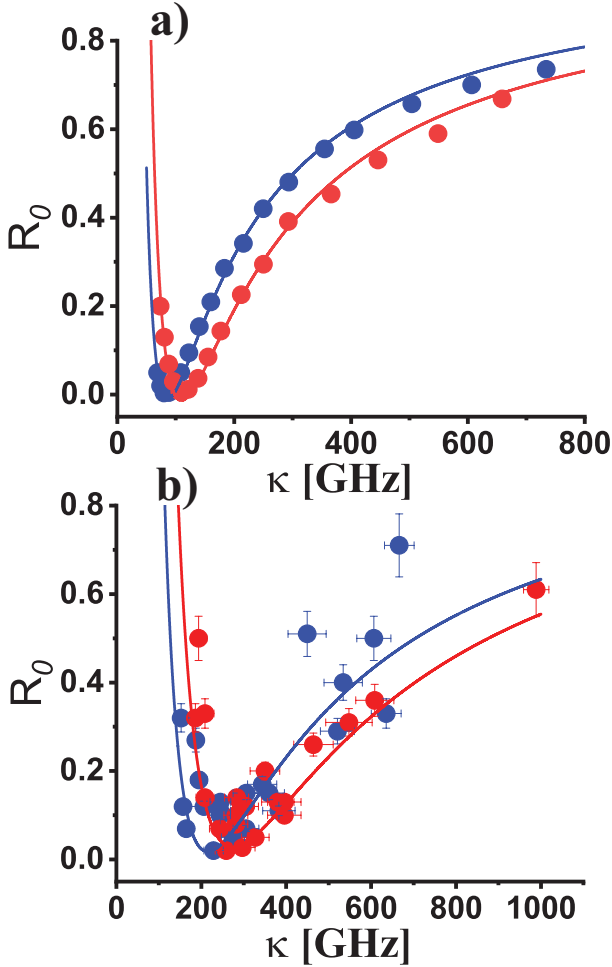}
\caption{{\label{Results2}} Simulated and the corresponding measured results for on-resonance reflection ($R_0$)-values. (a) and (b) show simulated and measured $R_0$-values as a function of total cavity linewidth ($\kappa$). Blue (red) dots correspond to $x$ ($y$)-mode. Blue (red) solid lines are fit to the data. }
\end{figure}

Reflection amplitude ($r$) for one-sided cavity can be formulated \cite{Milburn,Jam17} as
\begin{equation}
r= \frac{\frac{1}{2}(\kappa_{in}-\kappa_{sc})-i\Delta\omega}{\frac{1}{2}(\kappa_{in}+\kappa_{sc})+i\Delta\omega}, 
\end{equation}
where $\kappa_{in}$, $\kappa_{sc}$, and $\Delta\omega$=  $\omega-\omega_{0}$ are input coupling rate, scattering rate (intra-cavity loss rate), and detuning between the light source frequency ($\omega$) and cavity resonance frequency ($\omega_{0}$), respectively. $\kappa$= $\kappa_{in}+\kappa_{sc}$ is the total cavity linewidth. The reflectivity is $R$= $\mid$$r$$\mid^2$. On-resonance ($\Delta\omega$= 0) reflectivity ($R_{0}$) can be written as 
\begin{equation}
R_{0}= \mid\frac{\kappa_{in}-\kappa_{sc}}{\kappa}\mid^2= \mid1-\frac{2\kappa_{sc}}{\kappa}\mid^2
\end{equation}
The above equation implies that as $\kappa$-value decreases, $R_0$-value decreases in the over-coupling regime ($\kappa/2$>$\kappa_{sc}$), reaches zero at critical-coupling ($\kappa/2$=$\kappa_{sc}$=$\kappa_{in}$), and increases in the under-coupling regime ($\kappa/2$<$\kappa_{sc}$). 

Simulated and the corresponding measured coupling characteristics of the present cavity are summarized in Figs.~\ref{Results2}(a) and (b), respectively. Simulated $R_0$-values as a function of $\kappa$-values are plotted in Fig.~\ref{Results2}(a). Blue (red) dots correspond to $x$ ($y$)-mode. $N_{in}$-value is varied from $70$ to $240$ ($70$ to $220$) for $x$ ($y$)-mode. To infer scattering limited decay rate ($\kappa_{sc}$) of the formed cavity, we fit simulation predicted $R_0$-values with $\kappa_{sc}$-value as free parameter using Eq. (2). Fitted result for $x$ ($y$)-mode $R_0$-values is shown by the blue (red) solid line. For $x$ ($y$)-mode, obtained $\kappa_{sc}$-value is $43$ GHz ($54$ GHz). Obtained $\kappa_{sc}$-value for  $y$-mode is higher than that for $x$-mode due to a higher scattering loss of $y$-mode.

We measured cavity reflection spectra for $N_{in}$-value from $100$ to $190$ i.e.various $\kappa$-values. Measured $R_0$-values behavior is shown in Fig.~\ref{Results2}(b). Blue (red) dots correspond to $x$ ($y$)-mode. The origin of horizontal error bars is mainly due to the mounting uncertainty of ADMG onto ONF during experiments and the resolution of the spectrum analyzer used for measurements. The origin of vertical error bars is mainly due to power fluctuations of injected supercontinuum light source used for measurements. By increasing $N_{in}$-value i.e. decreasing $\kappa$-value, we clearly observe $R_0$-value transition from over- to- critical- and under-coupling regimes as predicted by simulations. One can readily see the expected behavior of $R_0$-values for $x$ ($y$)-mode as predicted by simulations. We estimate $\kappa_{sc}$-value by fitting Eq. (2) to the measured $R_0$-values. Blue (red) solid line is the fitted result for $x$ ($y$)-mode $R_0$-values. For $x$ ($y$)-mode, obtained $\kappa_{sc}$-value is $107$$\pm$8 GHz ($136$$\pm$10 GHz). Note that measured $\kappa_{sc}$-value for $y$-mode is higher than that for $x$-mode as predicted by simulations. 

Regarding measured $\kappa_{sc}$-value, deviation from the simulation predicted value is mainly due to the fabrication ambiguities in ONF diameters and ADMG parameters, leading to a higher scattering loss. Using $\kappa_{sc}$-value and effective cavity length ($l$) of $22$ $\mu$m, we estimate scattering limited cavity quality factor ($Q_{sc}$= $\frac{c}{\lambda_0\kappa_{sc}}$), cavity finesse ($\mathcal{F}_{sc}$= $\frac{c}{2l\kappa_{sc}}$), and one-pass power loss ($\mathcal{L}$= $1-e^{-2\pi\kappa_{sc}l/c}$).  Simulated and measured $Q_{sc}$, $\mathcal{F}_{sc}$, and $\mathcal{L}$-values are summarized in Table~\ref{Table2}. Superscripts denote polarizations. Note that the simulated $Q_{sc}$ and $\mathcal{F}_{sc}$ ($\mathcal{L}$)-values give an upper (lower) bound on cavity performance for both polarizations. Measured $\mathcal{L}$-value mainly indicates insertion loss due to ADMG as the ONF loss is much lower than 2\%. Simulated and measured results clearly demonstrate that ADMG does not introduce significant losses while forming a one-sided composite cavity on ONF. $Q_{sc}$, $\mathcal{F}_{sc}$, and $\mathcal{L}$-values can be improved by designing ADMG to produce an apodization index modulation\cite{Loncar,KaliPhC1}. It should be noted that high channeling efficiency can still be realized in the over-coupling regime with $\kappa$-value in the range $400$-$600$ GHz (see Fig.~\ref{Conceptual diagram}(c)). Therefore, the present cavity can be implemented for efficient unidirectional single photon generation using solid-state single quantum emitters. Note that unidirectional channeling of single photons can also be achieved based on chiral interactions\cite{Lodahl}, but it can only control the photon flux into one side. In contrast, the one-sided cavity method is to enhance the unidirectional channeling efficiency and Purcell enhancement to realize deterministic single-photon sources\cite{Haroche,Thompson, Fushman, Tiecke,Englund}.  
  
\begin{table}
\caption{{\label{Table2}}Simulated and measured performance parameters for the one-sided composite cavity on optical nanofiber.}
\centering
\begin{tabular}{|p{1.4cm}|p{0.7cm}|p{0.7cm}|p{0.7cm}|p{0.7cm}|p{0.7cm}|p{0.7cm}|p{0.7cm}|p{0.7cm}|}
\hline
 &\multicolumn{2}{c|}{$\kappa_{sc}[GHz]$}&\multicolumn{2}{c|}{$Q_{sc}$}&\multicolumn{2}{c|}{$\mathcal{F}_{sc}$}&\multicolumn{2}{c|}{$\mathcal{L} [\%]$}\\
\cline{1-9}
&\it{$\kappa_{sc}^{x}$}&\it{$\kappa_{sc}^{y}$}&\it{$Q_{sc}^{x}$}&\it{$Q_{sc}^{y}$}&\it{$\mathcal{F}_{sc}^{x}$}&\it{$\mathcal{F}_{sc}^{y}$}&\it{$\mathcal{L}^x$}&\it{$\mathcal{L}^y$}\\
\cline{1-9}
Simulation&43&54&10900&8680&155&124&2.0&2.5\\
\hline
Experiment&107&136&4478&3518&$62$&$49$&$4.9$&$6.2$\\
&$\pm$8 &$\pm$10 &$\pm$334 &$\pm$260&$\pm$4 &$\pm$3&$\pm$0.4&$\pm$0.5 \\
\hline
\end{tabular}
\end{table}

In summary, we have demonstrated a one-sided composite cavity on an ONF. Numerical and experimental results have clearly shown that the composite method can be extended to a one-sided cavity scheme without any significant scattering losses. The presently designed one-sided composite cavity can channel total spontaneous emission into one side of ONF. Thus, the designed one-sided composite cavity with a narrow bandwidth single quantum emitter may open new routes in nanofiber quantum photonics\cite{Mark17,Shafi,Shafi20,Neu}. The one-sided cavity can be advantageous for practical device design and applications in quantum information science.\\

This work was supported by the Japan Science and Technology Agency (JST) as one of the strategic innovation projects. RRY acknowledges the partial support by the Science Engineering Research Board (SERB), India (File No. SRG/2019/000989).\\

The authors declare that they have no conflict of interest.\\

The data that support the findings of this study are available from the corresponding author upon reasonable request.\\

\end{document}